\documentclass[final] {aipproc}
\layoutstyle{6x9}

\usepackage{graphicx}
\usepackage{amsmath}
\usepackage{amssymb}
\usepackage{bm}
\newcommand {\eqrefa}[1]{(\ref {#1})}
\renewcommand {\vec}[1] {\ensuremath{\boldsymbol{#1}}}
\newcommand {\R}{\ensuremath{\mathbb {R}}}

\begin{document}
\title{Retarded gravitation theory}
\classification{04.80.Cc, 04.50.Kd, 98.62.Dm, 45.20.D-, 02.30.Ks}
\author{C. K. Raju}{
address={School of Mathematical Sciences, Universiti Sains Malaysia, 11800 Penang, Malaysia}, 
altaddress={Current address: Albukhary International University, 05200 Alor Setar, Malaysia \\\url{ckr@ckraju.net}} 
}


\keywords {Retarded gravity, galactic rotation curves, flyby anomaly, functional differential equations, experimental tests of gravity}

\begin {abstract}
We propose a Lorentz-covariant theory of gravity, and explain its theoretical origins in the problem of time in Newtonian physics. 
In this retarded gravitation theory (RGT), the gravitational force depends upon both retarded position and velocity, 
and the equations of motion are time-asymmetric retarded functional differential equations. We explicitly solve these equations, under simplifying assumptions, for various NASA spacecraft. This shows that the differences from Newtonian gravity, though tiny within the solar system, are just appropriate to explain the flyby anomaly as a $\frac{v}{c}$ effect due to earth's rotation. The differences can, however, be large in the case of a spiral galaxy, and we show that the combined velocity drag from a large number of co-rotating stars enormously speeds up a test particle. Thus, the non-Newtonian behaviour of rotation curves in a spiral galaxy may be explained as being due to velocity drag rather than dark matter. RGT can also be tested in the laboratory. It necessitates a reappraisal of current laboratory methods of determining the Newtonian gravitational constant $G$. Since RGT makes no speculative assumptions, its refutation would have serious implications across physics.  
\end{abstract}

\maketitle

\section{Introduction}

The notion of time is fundamental to this reformulation of gravity. Though published long ago \cite{ckrtitcon}, my analysis of the notion of time is perhaps not so widely known, and this is briefly summarized below.

\subsection{The problem of equal intervals of time in Newtonian physics}

It is well known that Newtonian physics had a difficulty in defining equal intervals of time. One cannot bring back one hour from the past and put it side by side with one hour in the future and compare the two in the present to say by inspection that the two time intervals are equal. The equality of two time intervals is a matter of definition. Such a definition is needed to make sense of Newton's first law: ``uniform'' motion means a particle covers equal distances in equal times. This presupposes a theoretical clock: what is uniform motion according to a pendulum would not be uniform motion according to one's heart beats, or an atomic clock. Newtonian physics does not explicitly specify which clock to use.  For applications to planetary motion or ballistics, there are many clocks which one could use in practice. However, the lack of a theoretical definition of equal intervals of time became prominent when Newtonian physics had to be reconciled with Maxwellian electrodynamics. If we define equal intervals of time so as to make Newtonian physics valid, that makes Maxwellian electrodynamics invalid. 

Newton's predecessor, Isaac Barrow, was well aware of the need to specify a clock physically, and he said that those who were content to do physics with only a woolly idea of time were quacks. He defined equal intervals of time by saying that  equal causes take equal times to produce equal effects. On this definition, a sand glass measures equal intervals of time. Though crude, this was at least a physical definition. \cite{ckrSced} 

Newton, however, took a step backwards. In his \textit{Principia} he accepted that there was perhaps no physical clock which measured  equal intervals of time accurately. But he was unperturbed by this since he thought time was mathematical, meaning metaphysical. His well-known quote about ``absolute, true, and mathematical time'' flowing on ``without regard to anything external'' uses four adjectives to emphasize his belief in time as metaphysical and his further belief that this metaphysics was fundamental to his formulation of physics. That is, Newton thought it enough that equal intervals of time were known to God, for he believed God wrote the laws of Nature in the perfect language of mathematics. However, to do physics, human beings too need to know what equal intervals of time are. In  the absence of  a definition it is not meaningful to say that clock $A$ is more accurate than clock $B$. At best, one can say, as Poincar\'e eventually did, that one clock is more \textit{convenient} than another. 

\subsection{Calculus, the continuum, and the nature of time}

What is not so well known is \textit{why} Newton insisted on making time metaphysical (thus avoiding a physical definition of equal intervals of time). This was related to his perception of mathematics as perfect. Newton's second law needs $\frac{d}{dt}$, and he thought making $t$ metaphysical was the answer to Descartes' objection that the calculus was not rigorous, hence not mathematics. The intuitive idea of flow underlying Newton's fluxions was meant to make time infinitely divisible. While the idea of flowing time has been abandoned today, somewhat similar attitudes nevertheless persist: students are taught that a rigorous formulation of the calculus needs limits hence formal reals, $\mathbb{R}$, constructed through Dedekind cuts (no ``gaps''). Hence, if physics is formulated using differential equations, time must be like the real line. This procedure of allowing the mathematical understanding of calculus to dictate the physics of time is incorrect from the viewpoint of both physics and mathematics.

Mathematically, it is known (but not well known) that the continuum is \textit{not} essential to the calculus, which can be formulated in different ways. For example, one can do calculus in fields larger than \R. Since \R\ is the largest ordered field with the so-called Archimedean property, any ordered field, $F$, larger than \R\ must be non-Archimedean. That means there must be an element $x \in F$ such that $x > n$ for all natural numbers $n$. It would be appropriate to call such a number $x$ infinite. Since $F$ is a field, we must also have $0< 1/x < 1/n$, for all natural numbers $n$, so that $1/x$ is infinitesimal. Unlike non-standard analysis where infinities and infinitesimals enter only at an intermediate stage, directly using a non-Archimedean ordered field means the infinities and infinitesimals are ``permanent''. Limits in a non-Archimedean field are unique only up to infinitesimals. The field of rational functions is a simple example of such a non-Archimedean ordered field, and an elementary exposition of this field may be found in texts. \cite{Moise}

Historically speaking, the calculus developed in India, long before Newton, by using precisely this field of rational functions. Polynomials were called unexpressed numbers, so that rational functions were a natural counterpart of fractions, being unexpressed fractions. Of course, the infinite series of the calculus, and this way of handling convergence and limits by discarding infinitesimals was too sophisticated for European mathematicians then, since the practical mathematics of fractions had only just been introduced in the Jesuit mathematics syllabus  in the late 16th c. Hence, like Descartes, they expressed doubts about the meaning of an infinite sum, for they failed to understand the way infinite series of the Indian calculus were summed in Indian texts translated and imported into Europe by Jesuits based in Cochin. \cite{Cultfound}

Similarly, there are philosophies of mathematics other than formalism, and one can use zeroism \cite{Cultfound} to do calculus rigorously over floating point numbers. Floats  are not a field (the associative law for addition fails \cite{Hawaii}), and they are a finite set, hence smaller than \R. Thus, the continuum is not essential to the calculus, which can be done over mathematical structures both larger than \R\ or smaller than \R, so the text-book way to do calculus is not good reason to declare time to be like the continuum. 

Indeed, the use of the continuum presents difficulties. Thus, according to the calculus of college texts, discontinuous functions cannot be differentiated. However, the need to do so arises, for example, with shock waves. The Schwartz  derivative does \textit{not} generalize the college-text derivative, and if physics is to be kept refutable, one cannot just hop from one definition of the derivative to another. In fact, both definitions of the derivative are inadequate to deal with discontinuous solutions of nonlinear partial differential equations, the one because discontinuous functions cannot be differentiated, and the other because the derivatives of discontinuous functions, such as the Dirac $\delta$-function, cannot be readily  multiplied. \cite{distrib} However, I will not go further into those issues, here. 

Briefly, what is today taught as the authoritative way to do calculus is hardly the only way or the best way to do it. Therefore, \textit{theories of the calculus ought not to decide the nature of time in physics, as Newton did}. 

One can also object to Newton's procedure on the grounds that it is inappropriate to allow metaphysical beliefs to dictate something as fundamental to physics as the nature of time. Time may be like the real line, or it may be discrete or structured. But, whatever the nature of time, this should be decided by physics, not by mathematics. 

\subsection{Special relativity and FDEs}

Anyway, we know how the issue of defining equal intervals of time was historically resolved. Poincar\'e defined equal intervals of time by postulating that the speed of light, $c$, is constant. This allows equal intervals of time to be defined using a photon bouncing between parallel mirrors. This postulate led directly to the special theory of relativity. From this point of view, special relativity originated as a solution to the subtle problem of defining equal intervals of time in Newtonian physics. Poincar\'e selected this definition on the explicit grounds of convenience. In the absence of a theoretical definition of a proper clock, the Michelson-Morley experiment and the later series of counter-experiments by Miller could have neither established nor refuted the constancy of the speed of light. \cite{syngeSR, ckr3a}

The postulated constancy of $c$ implies that interactions cannot propagate faster than $c$ (in either direction in time). This has an important mathematical consequence not mentioned in texts on special relativity. \textit{The delay in propagating interactions means that Newtonian physics ought to be reformulated using delay differential equations}(or, more generally, functional differential equations [FDEs]). 

Poincar\'e understood this point. Einstein, to whom the theory of relativity is credited, did not understand this. Till the end of his life, he kept trying to get rid of the delay by using the Madhava (``Taylor'') series expansion to approximate delay differential equations by ordinary differential equations (ODEs). We have to understand that this is a serious mathematical mistake, for solutions of FDEs exhibit qualitative behavior impossible for ODEs. This matter has been discussed in detail in my earlier books and I will only summarize those examples here. 

Thus, consider the FDE $\dot y (t) = -2 y(t) + y(t-r)$, where $r > 0$ is a small constant. It may be shown (using the characteristic quasi-polynomial) that every solution of this FDE is bounded and tends to 0 as $t \to \infty$. However, if we ``Taylor'' expand the term $y(t-r)$ and truncate after two terms, we obtain the linear ODE with constant coefficients $\dot y(t) = -2 y(t) + [y(t) - r \dot y(t) - \frac{1}{2} r^2 \ddot y(t)]$ which admits exponentially increasing solutions $y(t) = e^{\alpha t}$, with $\alpha > 0$, no matter how small $r$ is, so long as $r > 0$. Thus, ``Taylor'' expanding to reduce an FDE to an ODE leads, in general, to qualitatively incorrect behaviour.

The FDE $\dot y = f(t,~y(t),~y(t-\tau))$  is called retarded if $\tau > 0$. To solve this FDE for $t > 0$, it is not enough to specify only the initial data $y(0)$, as with ODEs. For example, the equation $\dot y = y(t-\pi/2)$ has  $\sin t$ and $\cos t$ as obvious solutions. Since the equation is linear $a \sin t + b \cos t$ is a solution for arbitrary constants $a$ and $b$. Both $a$ and $b$ cannot be fixed from a single number $y(0)$. In general, one must prescribe \textit{past} data rather than initial data. That is, the entire function $y(t)$ must be specified on a past interval $[-r, 0]$, where $r > 0$ is just $\tau$ when the latter is constant. In the state-dependent case (when $\tau \equiv \tau (y)$), knowledge of the entire past $(-\infty, 0]$ may be needed to calculate solutions for \textit{all} future time. With past data appropriately specified, it is possible to prove the existence of solutions and also to compute them numerically.   

But, from the viewpoint of physics there is a further problem. \textit{How} should  past data be prescribed? This problem also crops up in classical electrodynamics. If we use the full electrodynamic force (not just the static Coulomb force) the equations of motion of a charged particle are FDEs. In this case the FDEs arise through the coupling of the Heaviside-Lorentz force law (ODE) with the Maxwellian partial differential equations (PDE). For some reason, for over a century, this problem was not addressed in the physics literature, and the FDEs of the simplest 2-body problem of classical electrodynamics remained unsolved in any serious physical context, until I first solved them in 2004. \cite{ckrem2bp} As explained in that paper, once it is noticed that FDEs are equivalent to a coupled system of ODEs + PDEs then prescribing past data, or the past world-lines of particles, is seen to be equivalent to prescribing fields on a Cauchy hypersurface needed to solve the Maxwellian PDE.

\begin{figure}
\centering
\includegraphics[width=8cm]{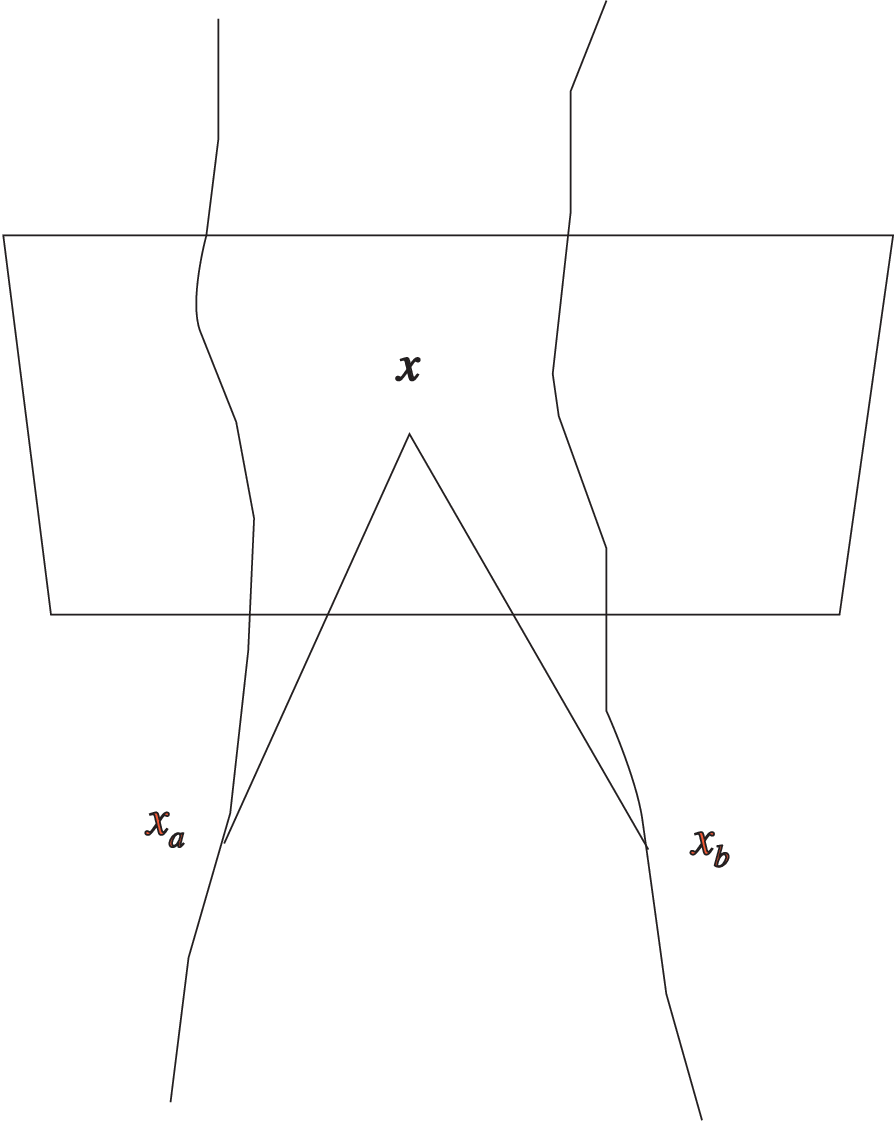}
\caption{The figure shows how prescribing initial data on a Cauchy hypersurface (for fields) is equivalent to prescribing past data on the world lines (for particles). Assuming retarded propagators, and only two particles, the field at point $x$ on the hypersurface is determined by the world lines of the two particles at the points $x_a$ and $x_b$ where the backward null cone from $x$ meets the two world lines. As $x$ runs over the hypersurface, the points $x_a$ and $x_b$ would, in general, cover the entire past world-lines of the two particles.}
\label{pastdata}
\end{figure}

\subsection{FDEs and irreversibility}

The widespread neglect of FDEs, by physicists, is a pity because FDEs offer several advantages from the viewpoint of physics.  Thus, Newtonian physics uses ODEs which are time-reversible. This is contrary to the everyday experience that time is fundamentally irreversible. 

This conflict between experience and theoretical expectation may be seen simply as experimental refutation of Newtonian physics. However, physicists are taught to see this instead as a problem of thermodynamics: the problem of reconciling irreversible entropy increase with reversible physics. Physics texts insist that coarse graining and the Ehrenfest model may be used to resolve the recurrence and reversibility paradoxes, though there are numerous objections to these supposed refutations. \cite{ckrThermo} 

For example, the recurrence paradox is resolved by claiming large recurrence times. However, the recurrence time cannot be calculated without a host of assumptions. A key assumption is ergodicity or mixing. Without this assumption, the recurrence time need not be large. However, no one has the foggiest idea of the physical characteristics of  an ergodic system, and no one can say whether the cosmos is ergodic. Underlying the assumption of ergodicity is the assumption of stochastic, Markovian evolution which is taken, not as the real physical model of time evolution, but as the apparent reality. 

That is, the text-book way of resolving the ``paradoxes of thermodynamics'' amounts to this:  the most fundamental aspect of everyday experience, irreversible aging, is declared an illusion. This is dangerous, for any experimental fact, contrary to any theory,  can always be ``explained'' away by piling on the hypotheses. If this ``explanation'' of irreversibility is intended seriously, it should also explain how to reverse aging. As far as I know, no one has done that till now.

But even this way out is not available in general relativity theory (GRT). Thus, the recurrence theorem extends to any kind of time-symmetric and deterministic evolution \cite{ckrThermo}.  This includes the case of geodesic flow on a manifold. However, the calculation of recurrence times cannot be so extended. Indeed, the very notions of Lorentz-invariant probability measure and Lorentz-invariant Markov process present deep difficulties, and GRT has only very limited analogues to classical statistical mechanics.

\textit{The alternative is to accept the empirical fact of irreversibility and to reformulate dynamics in an irreversible way ab initio. This is precisely what we do, using retarded FDEs.} A retarded FDE can be solved forward in time, but not backward. That is, we cannot reconstruct the past from the future, though we can predict the future from the past. 

\begin{figure}[h]
	\centering
		\includegraphics [width=8 cm]{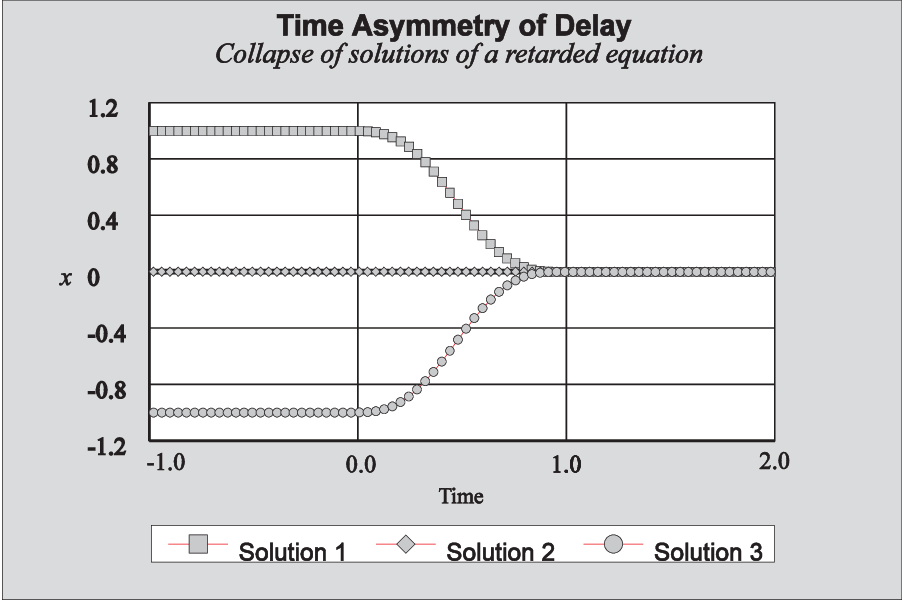} 
	\caption{Three solutions of a retarded FDE which collapse towards the future}
\label{retarded}
\end{figure}

For example, consider the retarded FDE $\dot y(t) = b(t) y(t-1)$, where $b(t)$ is a continuous function which vanishes outside [0,~1], and satisfies $\int b(t)~dt = -1$. For example, For example, 
\begin{equation}
b(t) = \left  \{  
\begin{array}{r @{\quad : \quad} l}
0 & t \le 0 \\
-1 + \cos 2 \pi t & 0 \le t \le 1 . \\
0 & t \ge 1
\end{array}
\right .
\end{equation}
Three solutions of this equation are shown in Fig.~\ref{retarded}, using different past data prescribed on $[-1, ~0]$. Since the solutions are identical for $t > 1$, past values cannot be reconstructed from knowledge of the solution for future times,  $t > 1$. 

According to the information-theoretic understanding of entropy, \cite{ckrThermo} this loss of information towards the future, is exactly the same as the claim that entropy increases towards the future: we have more information about the past than the future. Further, as the above figure shows, retarded FDEs are not only irreversible, they exhibit phase collapse. The proof of the recurrence theorem fails if volume in phase space is not conserved. Consequently, both reversibility and recurrence paradox are cleanly resolved by retarded FDEs.

\subsection{Retarded gravitation and flyby anomaly}

These retarded FDEs arise most naturally, if the special relativistic correction of Newtonian physics is carried to its logical conclusion. While electrodynamics already has retarded FDEs, the Newtonian gravitational force law too needs to be changed into a retarded force law. Once we have defined equal intervals of time by postulating the constancy of $c$, gravitational forces too cannot any longer propagate at a speed faster than $c$. We carry out this modification of the gravitational force law in \S 2 below to construct the retarded gravitation theory (RGT). 

As one might expect, RGT differs very marginally from Newtonian gravity at the level of the laboratory or the solar system. To show this we explicitly solve the equations of RGT for the earth flyby of various NASA spacecraft. 

Between 1990 and 2005, six NASA spacecraft flew by earth, using the technique of earth gravity assist, to either gain or lose heliocentric orbital energy. Anderson et al.  \cite{Anderson2008}
reported that anomalous orbital energy changes were observed. The anomalies were extremely small, corresponding to an unexplained velocity difference of the order of a few mm/s (while the spacecraft velocity at perigee is of the order of 10 km s$^{-1}$) but these were well beyond systematic experimental error, which ranged from 0.01 mm s$^{-1}$ to 1 mm s$^{-1}$. The causes of the anomalies could not be explained despite a careful audit and consideration of various possible factors including general relativistic effects. \cite{Lammerzahl} 

Further, Anderson et al. found an empirical formula which fitted all six flybys:
\begin{equation} 
\frac{\Delta V_\infty}{V_\infty} = K ( \cos \delta_i  - \cos \delta_o ) ,
\label{Anderson formula}
\end{equation}
where $\Delta V_\infty$ was the difference between the incoming and outgoing asymptotic velocity in a \textit{geocentric} frame (conceptually the hyperbolic excess velocity at infinity of an osculating Keplerian trajectory, which difference ought to have been zero on the Newtonian theory) and $\delta_i$ and $\delta_o$ were the declinations of the incoming and outgoing asymptotic velocity vectors. The constant $K$ was expressed in terms of the Earth's angular rotational velocity
$\omega_E$ of $7.292115 \times 10^{-5}$ rad/s, its mean radius $R_E$ of 6371 km and the speed of light $c$ by
$K = \frac{2\omega_E R_E}{c} = 3.099 \times 10^{-6}$. Combined with the dependence on declination in \eqref{Anderson formula}, this suggested that the observed anomalous effect in the flybys depended systematically on the rotation of the earth.

Using RGT to solve for the motion of these spacecraft brings out two things. First, \textit{the difference between RGT and Newtonian gravity is tiny} and of the order $\frac{v}{c}$, where $v$ is the rotational velocity of the earth. 
Secondly, \textit{this $\frac{v}{c}$ effect arises naturally in RGT, but cannot be explained either by Newtonian gravity or GRT}.  
It is, of course, possible to ``save'' the existing theories by appealing to a variety of confounding factors to claim that the observed anomaly is somehow not ``real''. However, according to RGT the anomaly is real and will be again observed if the observations are repeated more carefully, and after eliminating confounding factors. To put it simply, if the effect is real, Newtonian gravitation and GRT are refuted, and if not, RGT is refuted. Of course, whether or not the effect is real should be decided by experiments and not merely the opinions of ``authorized'' physicists. Similar experiments can, in principle, even be carried out in the laboratory as we explain in \S 4. 

\subsection{RGT, GRT and galactic rotation curves}

By some quirk of history, after Poincar\'e, the modification of Newtonian gravitation, to make it Lorentz covariant, was never pursued. Instead, we had GRT, which goes a step further and dispenses with the concept of force itself. Now, I happen to agree that the notion of force is an artificial notion which is better eliminated. However, the way this was done by GRT, it ended up losing also the simplifying concept of a point mass.

As already noted above, the continuum description of matter in GRT is not one which is statistically related to an underlying assembly of discrete molecules as in classical mechanics. \cite{syngeGR} Further, one \textit{sees} planets and stars as discrete blobs of matter, but their representation in GRT is not straightforward. As also indicated above, the related problem of junction conditions in GRT \cite{ckrJunction} is open to dispute, just because the PDEs of GRT can be (and have been \cite{distrib}) interpreted in different ways in the presence of discontinuities.   

All this makes it very difficult to do the many-body problem with GRT. Naturally, given a complicated many-body problem, there is a relapse to simpler Newtonian physics, in practice. Newtonian gravity is what is used to solve the many-body problem in a galaxy. 

However, it has long been known that the observed rotation of stars in spiral galaxies belies Newtonian expectations. If $v(r)$ is the velocity of a star at (average) distance $r$ from the galactic center, then at the edge of the galaxy we should have $v(r) \propto r^{-\frac{1}{2}}$ on Newtonian gravitation. However, instead of declining as expected, $v(r)$ is observed to \textit{increase} with $r$,  eventually becoming \textit{constant}  \cite{RubinA}.  

From the present point of view, the above departure from Newtonian dynamics is yet another indication of the empirical failure of Newtonian physics, beyond planetary orbits, and invisible dark matter is yet another hypotheses being used to ``save'' the theory from refutation. 

While the hypothesis of dark matter has some plausibility, the mere existence of dark matter does not explain why $v(r)$ becomes constant. Accordingly, the further hypothesis is advanced that the dark matter is distributed like a halo, with its density reaching a peak where the density of luminous matter thins out to zero. Since the existence of the dark matter is inferred solely by its gravitational effects, which (regardless of its composition) are assumed to be identical to those of luminous matter, it is hard to understand why luminous and dark matter should be distributed so very differently. Thus, dark matter halos seem an artificial hypothesis invented just to save the Newtonian theory from manifest failure. 

RGT offers a simple and natural explanation for the seemingly peculiar behaviour of rotation curves in a spiral galaxy, without any additional hypothesis. The explanation is that a spiral galaxy, unlike the planetary system, contains a large number of co-rotating stars. The combined velocity drag accelerates the rotation speed of outer layers, and we see this in more detail in \S 3. In general, RGT offers a simple way of doing many-body dynamics, in a Lorentz covariant way, which improves on Newtonian theory, while this simplicity is not available in GRT. 

\subsection{RGT and other modifications of gravity}

We reiterate that the differences between RGT and Newtonian gravitation can, in principle, be tested in the laboratory. 
This makes RGT quite different also from modified Newtonian dynamics or  MOND  \cite{MilgromA} which supposes (on phenomenological grounds) that the gravitational force law somehow changes, but only at the level of the galaxy. In contrast, RGT changes the Newtonian gravitation force law at all levels, in one and the same way.  Many other modifications of gravity have been proposed, but none is open to direct empirical test in the laboratory, like RGT. 

Indeed, we emphasize that, unlike all earlier modifications of gravity, \textit{RGT is not a speculative theory}, for we have assumed nothing beyond Lorentz covariance. Further, we have  explained why even Lorentz covariance is essential in order to be able to measure time. Hence empirical failure of the theory would have serious repercussions across physics. RGT also calls into question the assumptions underlying experiments to measure the Newtonian gravitational constant $G$ (\S 4).

The need to modify Newtonian gravitation to make it Lorentz covariant was noticed long ago by Poincar\'e \cite{Poincare1906}. However, his starting point was different (a theory of the electron), and he explored various possible mathematical expressions for the gravitational 4-force without, 
fixing on a definite expression for it. He rightly emphasized that fixing a definite expression would be premature in the absence of empirical data. He obviously never explored the consequences of Lorentz covariant gravity for the rotation curves of galaxies, or the trajectories of spacecraft, which consequences were not known in his time. It is the availability of this empirical information which has helped us to fix a specific form of the new gravitational force law.

\section{The retarded gravitational force}

We start with a reference frame in which the test particle (``attracted body'') is a mass point (at rest)
at the origin. 
The  ``attracting body'' is located at the retarded position described by the 4-vector $X = (ct, \vec{x}) $ and moving with a 4-velocity $V = \gamma_v (c, \vec{v})$, both at \textit{retarded} time $t = -\frac{r}{c}$. 
Here, $\vec{x} = (x, y, z)$,  $r = \sqrt{x^2 + y^2 + z^2}$, and $\gamma_v = (1- \frac{v^2}{c^2})^{-\frac{1}{2}}$ is the Lorentz factor. 
 
Let $F = (T, \vec{f})$ 
be the 4-force experienced by the attracted body. This 4-vector transforms in the same way as the 4-vectors $X$  and $V$, so 
we take it to be
given by a linear combination
\begin{equation} 
F = a X + b V ,
\label{basic}
\end{equation}
where $a$, and $b$ 
are Lorentz invariants to be determined. Since  $a$ and $b$ are Lorentz \textit{invariant}, the expression \eqrefa{basic} for the 4-force $F$ would be Lorentz covariant, as required.

For the case where the attracting body is also at rest ($\vec{v}= 0$), we require that the 3-force  must approximately agree with the Newtonian gravitational force $\vec{f} =  k(\frac{x} {r^3}, \frac{y}{r^3}, \frac{z}{r^3})$, where $k = Gm_0 m_1$, the two (rest) masses are $m_0$ and $m_1$, and $G$ is the Newtonian gravitational constant. 
(Note that the sign conventions we are using are the opposite of the usual ones, since the ``attracting body'' is at $X$, and the force is in the direction of its retarded position.) Therefore,  $a \approx \frac{k}{r^3}$. This suggests that $a = -\frac{kc^3}{a_1^3}$ where $a_1$ is  the Lorentz invariant quantity  $a_1 = X.V = \gamma_v (c^2 t - \vec{x}.\vec{v})$, which 
equals $-cr$ when $\vec{v}=0$, and 
approximately equals $-cr$ when $v= || \vec{v}||$ is small compared to $c$. That is, 

\begin{equation}
	a = -\frac{kc^3}{(X.V)^3} \approx \frac{k}{r^3} .
	\label{a}
\end{equation}
 
We now use the fact that the components of the 4-force are not independent, but must satisfy  \cite{SyngeandGriffith}
\begin{equation}
F.U = 0,
\label{fdotu}
\end{equation}
where $U = \gamma_u (c, \vec{u})$ 
is the 4-velocity of the particle on which the force acts.
This comes about simply since the revised form of the equations of motion is now
\begin{equation}
m_0 \frac{d^2 Y}{  d s^2} = F ,
\label{eom}
\end{equation}
where $m_0$ is the rest mass and $s$ is proper time along the world line, $Y(s)$, of the ``attracted particle''. Since the 4-force $F$ is parallel to the 4-acceleration of the particle on which it acts, it must be perpendicular to its 4-velocity $U$ (which is a vector of constant norm). Accordingly, taking the dot product of $U$ with both sides of \eqrefa{basic}, we obtain
\begin{equation}
0 = a(X.U) + b (V.U) .
\label{b}
\end{equation}
 
Now the dot products $X.U$ and $V.U$ are scalars, or Lorentz invariants, and the Lorentz invariant $a$ is already determined. Hence, \eqrefa{b} 
determines $b$ as a Lorentz invariant. Explicitly, 
\begin{equation}
	b = - \frac{a (X.U)}{(V.U)} \approx \frac{k}{cr^2} .
	\label{b1}
\end{equation}

Note that we would not have been able to satisfy the requirement  \eqrefa{fdotu} had we already set $b = 0$ to begin with. This shows that the Lorentz covariant force we seek \textit{cannot} be purely position dependent.
Substituting these values of $a$ and $b$ in \eqrefa{basic}, the force in RGT is explicitly given by 
\begin{equation}
F = -\frac{kc^3}{(X.V)^3} X + \frac{kc^3}{(X.V)^3}\frac{(X.U)}{(V.U)} V .
\label{explicit}
\end{equation}
Since the equations of motion \eqrefa{eom}, and the expression for the force \eqrefa{explicit} are Lorentz covariant, we can use these expressions in any Galilean frame, and are not tied to any special frame. Note, however, that RGT, unlike GRT, is restricted to Galilean frames. 

Now, the velocities of spacecraft are typically a few kilometers per second while the rotational velocities of stars in galaxies are in the range of a few hundred kilometers per second. Both are small compared to the speed of light. Accordingly, for these cases, 
in a frame in which the earth center or galactic center is at rest,
we can use the approximate expressions for $a$ and $b$ given in \eqrefa{a} and \eqrefa{b1}. This leads to
\begin{equation}
F \approx \frac{k}{r^2} \left (\frac{X}{r} + \frac{V}{c} ,
\right ) ,
\label{approx}
\end{equation}
which simple form exhibits the departure from Newtonian gravitation more clearly for non-relativistic velocities.

The new force has a component in the direction of the velocity of the attracting body. Given a large number of stars rotating in a common direction, it is evident that this new velocity component of the gravitational force will generate a ``velocity drag'' which will systematically speed up stars. For a typical galaxy, the number of stars is a large number ranging from $10^9$ to $10^{11}$. In a spiral galaxy,  because the stars are co-rotating, the velocity-drag will typically scale with the number of bodies involved. How much would it amount to?

\section{Solutions}

To show how to calculate the velocity drag, we now solve an $n$-body problem. 
This involves the solution of a system of retarded FDEs, for which we need to prescribe the \textit{past history} of the particles. A general procedure for doing this in physics was already laid down earlier \cite{ckrem2bp},
and a similar procedure can be followed even in the case of stiff (retarded) FDEs \cite{PaTomu}.

However, our immediate purpose is three fold. First, we need to show that for the cases of ballistics and planetary orbits, RGT differs very little from Newtonian gravitation. That is, we seek to identify the circumstances in which, despite the qualitative difference between FDEs and ODEs, the  solutions of the one are approximated by solutions of the other. Second, we seek to identify those differences between RGT and Newtonian gravity which cannot be explained by the latter (or even by frame drag in GRT). This is the case of the spacecraft motion being affected by the rotation of the earth. Third, we show that in the circumstances which prevail in a spiral galaxy, the differences between RGT and Newtonian gravitation are substantial. 

\subsection{Reducing FDEs to ODEs in the one-body case} 

To this end, consider a simplified model galaxy as a large central mass, surrounded by a ring of $n$ particles rotating with constant velocity larger than would obtain with Newtonian gravitation. Consider now an $(n+1)$st body or a test particle initially moving with the velocity prescribed by Newtonian gravitation, and below the velocity of the ring. On Newtonian theory, the rotational velocity of the test particle should stay unchanged. How would the velocity of this test particle change with RGT?

In the above scenario, we may reasonably suppose that the test particle has a negligible effect on the motion of the other $n$ bodies (namely the galaxy). That is, instead of prescribing just the \textit{past} history of the $n$ bodies, as required for the solution of FDEs, we may regard the \textit{entire} world lines of the $n$ bodies as given. In this situation, where the motion of all other particles is fixed, the equations of motion for the $(n+1)$st body become just ODEs. Exactly the same simplification can be used with spacecraft where the tiny effect of the spacecraft on the earth can be ignored. That is the FDEs of RGT reduce to ODEs when we are mainly concerned with the motion of only one small body, and can neglect its effect on the other bodies. (Note, however, that this reduction to ODEs would no longer be valid, if two comparable bodies are involved. Thus, retardation does not automatically make solutions of the 2-body problem unstable as it would in Newtonian physics.)

Using the approximation \eqrefa{approx}, and neglecting the $\gamma$ factors (for this case), the equations of motion can be rewritten in 3-vector form as 
\begin{eqnarray}
\vec{u} &= \frac {  d\vec{y}}{  d t} \label{trivial}\\	
\frac{  d \vec{u}}{  d t} &= \sum_{i=1} ^ n \frac{Gm_i} {r_i^2} \left  ( \frac{\vec{x_i}}{r_i}  + \frac{\vec{v_i}}{c} 
\right ) . \label{ode} 
\end{eqnarray}
Here $\vec{y}$ is the position vector of the test particle, $\vec{x_i}$ is the retarded relative displacement vector of the $i$~th  particle (so its absolute retarded position is $\vec{z_i} = \vec{y}+\vec{x_i}$). Thus, $(ct_i, \vec{z_i})$ is the point where the backward null cone from $(ct, \vec{y})$ meets the world line of the $i$~th particle. The sum on the right hand side of \eqrefa{ode} is over all $n$ attracting masses $m_i$, using their retarded distances $r_i$, and velocities  $\vec{v}_i$ at the retarded times $t_i$. For this preliminary calculation, we make a further simplifying assumption: given the symmetry of both models (of the galaxy and the earth) we expect that the sum over retarded quantities will be only slightly different from the sum over instantaneous quantities, which we accordingly use. We numerically solve this problem for the two cases, using Hairer et al.'s \textsc{dopri} program for ODEs. 

To do so, we need appropriate units \cite{ckrem2bp}
so that the numbers involved are neither too small nor too large.

\subsection{The solution for the galaxy}

For the galaxy, we solved \eqrefa{ode} using as units of length 1 kpc = $3.08568025 \times 10^{19}$ m, time $10^{14}$ s $\approx$ 10 million years, mass $1.98892 \times 10^{35}$ kg $\approx 10^5$ solar masses. In these units, we have the speed of light $c = 9.715603488080141 \times 10^2$, and $G = 4.5182366574577775 \times 10^{-6}$

For our model galaxy, we assumed a large (static) central mass ($1.5 \times 10^5$ mass units), surrounded by a ring of radius 12 kpc, consisting of 10000 particles, of 1 mass unit each,  rotating with an  angular velocity which is twice what one would expect on Newtonian considerations. Namely, $\omega_{\rm{ring}} = 3.96 \times 10^{-2}$, corresponding to a linear velocity of around 
146.6 km\, s$^{-1}$. 
The test particle is at 12.2 kpc, and initially rotating with an angular velocity expected on Newtonian considerations,  namely, $\omega_{\rm{particle}} 
\approx 1.94 \times 10^{-2}$
corresponding to a linear velocity of around 72.9 km\, s$^{-1}$. 
The exact mass of the test particle (assumed small) is irrelevant, and does not appear in the equations  \eqrefa{ode}. 

The solution is shown in the Figure~\ref{Fig1} below. The velocity of the test particle increases and it escapes. This is impossible with Newtonian gravitation.

The velocity effect persists \textit{inside} the ring where the Newtonian force due to the ring cancels, due to symmetry.
The sharpness of the velocity increase in the above figure is due to this: modeling the ring as infinitesimally thin, makes the transition from one regime to another abrupt. The velocity of the test particle decays at infinity as expected on Newtonian considerations but remains persistently larger, and the difference asymptotically approaches a constant. 

\begin{figure}[h]
	\centering
		\includegraphics [width=8 cm]{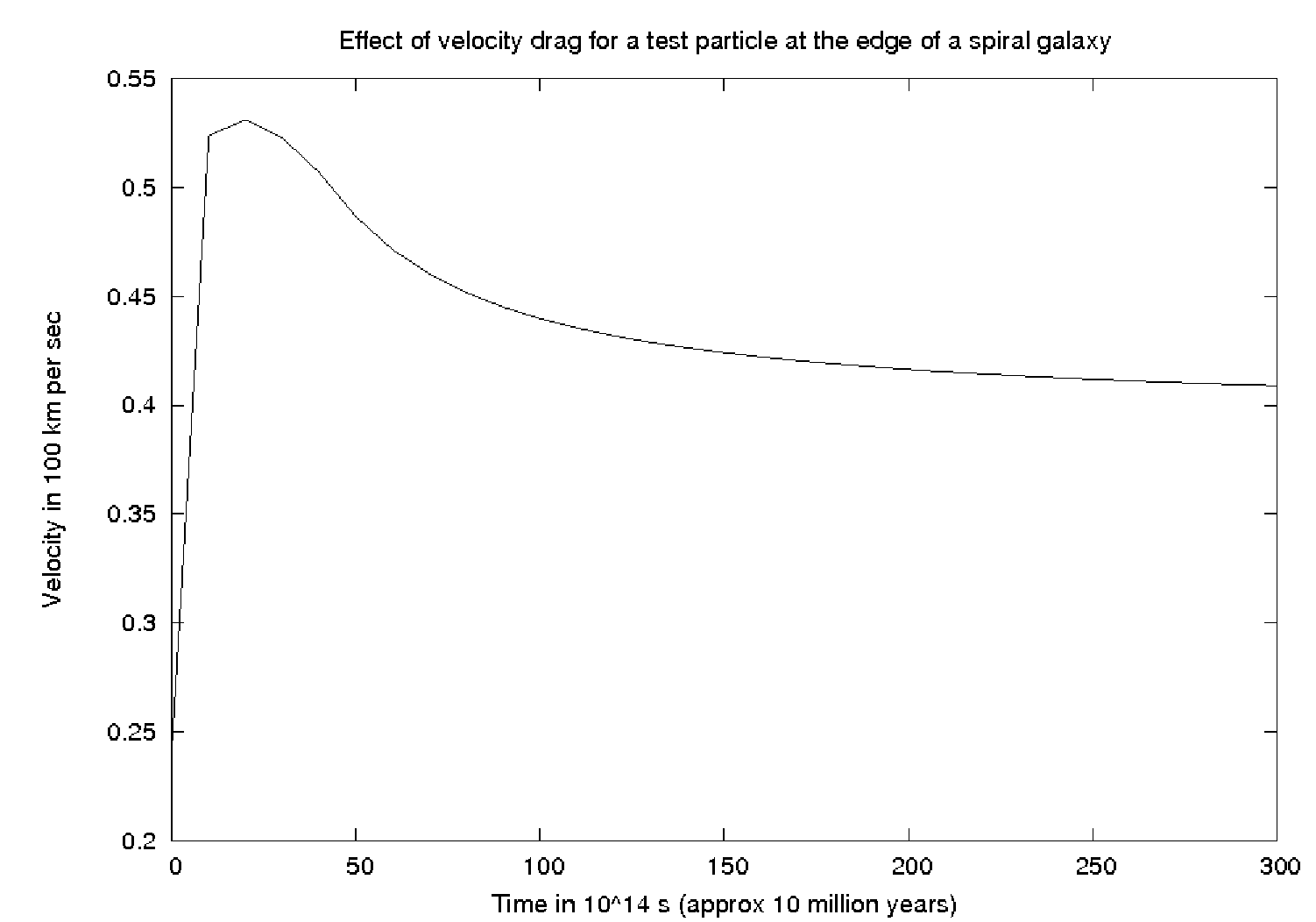} 
	\caption{\textbf{The velocity effect of retarded gravitation:} The velocity of a test particle in our model galaxy increases due to velocity drag, and the particle escapes.
The plot is velocity vs time. Time units are approximately 10 million years, while length units are 1 kpc. 
A central mass of $1.5 \times 10^{10}$ solar masses is surrounded by 10,000 particles (each of $10^5$ solar masses) in a rotating ring of radius 12 kpc. The test particle is initially in Newtonian equilibrium at 12.2 kpc.
}
\label{Fig1}
\end{figure}

Thus, with RGT, a large number of co-rotating particles leads to a systematic increase in the rotation velocity. Hence, in a spiral galaxy, a significantly higher-than-Newtonian rotational velocity due to velocity drag is to be expected on this theory. 
In the next refinement, suppose we model a spiral galaxy as a series of concentric rings. A slightly faster than Newtonian rotational velocity for an inner ring, would speed up the next outer ring and so on. A constant velocity at the edge of the galaxy, where the rings thin out, is a very plausible equilibrium situation with this model.

The solution shown above is clearly not the only one possible. There are two limiting cases: (a)~as we increase the number of rotating particles, the test particle is greatly speeded-up and escapes, or (b)~as we increase the central mass, so that the the Newtonian term in \eqrefa{ode} dominates, the test particle stays bound, and its velocity oscillates. 
The exact velocity at which the ring rotates is not critical. 
However, as the velocity difference between the ring and test-particle decreases the velocity dependent term becomes small, so the problem becomes numerically stiff and a different method of solution is needed. 

Naturally, we need to carry out calculations with more realistic models of spiral galaxies. For this, the above calculation provides proof-of-concept, for it shows that, contrary to what one might expect, the solution of a many-body problem with a Lorentz covariant theory can differ significantly from Newtonian gravity even at non-relativistic velocities, under the circumstances prevalent in a spiral galaxy.

\subsection{The solution for spacecraft}

For the case of a spacecraft, our computational units are length = 1 km, mass = $10^{18}$ kg, and time = 100~s. 
In these units, the spacecraft asymptotic velocities are typically in the range of hundreds of km per hundred seconds, and the effect we are looking for is in the range of mm/s or 0.0001 km per hundred seconds. As this is a first (``proof of concept'') computation, rather than a real-time engineering computation, to further simplify matters, and reduce computational complexity, we again used instantaneous distances instead of retarded distances. (Retarded distances could, of course, be used as was done earlier by this author in the electrodynamic case.)  

We use a simplified model of the earth as a perfect sphere of radius $R_E$, of uniform density, and rotating with a constant angular velocity $\omega_E$; this is discretized as a system of $n$ point masses whose (rotational) motions are prescribed in advance, where $n$ is a large number. (Some care is needed in this discretization. If we take a mesh which is uniform in spherical polar coordinates $r$, $\theta$, $\phi$, then, to ensure a uniform mass density in the $z$ direction, we must allow the masses at each mesh point to vary with the polar angle, to compensate for the smaller azimuthal circles at higher polar angles.) Doubtless this model can be improved, allowing for variations in density, oblateness and known gravitational anomalies etc. But, since our immediate interest is only to explain the physics of the gravitational velocity effect in RGT and indicate how it can be calculated, we postpone such refinements.

To solve for the spacecraft motion, we need initial data for the system of six ODEs \eqref{ode}, namely the initial position and velocity vectors for the spacecraft. Ideally, these should be obtained from direct observation. For our purposes, we use the ephemerides provided online by NASA's HORIZONS web interface (\url{http://ssd.jpl.nasa.gov/horizons.cgi}) which supplies the estimated position and velocity vectors in a geocentric inertial frame using the Earth mean-equator and equinox-of-reference-epoch (J2000.0). Ideally we should transform these coordinates to geocentric equatorial inertial coordinates for the epoch-of-date  to take into account variations due to the precession of the earth's rotation axis. 

Since the anomalous effect is tiny, to bring out the consequences of a velocity-dependent gravitational force,  and to diminish the confounding effects of artefacts of the modelling process, and simplifying assumptions, we also solved for the motion of the spacecraft with the same initial data, but using just the Newtonian force. 

\begin{figure}[!ht]
\centering
 \includegraphics[width=7 cm]{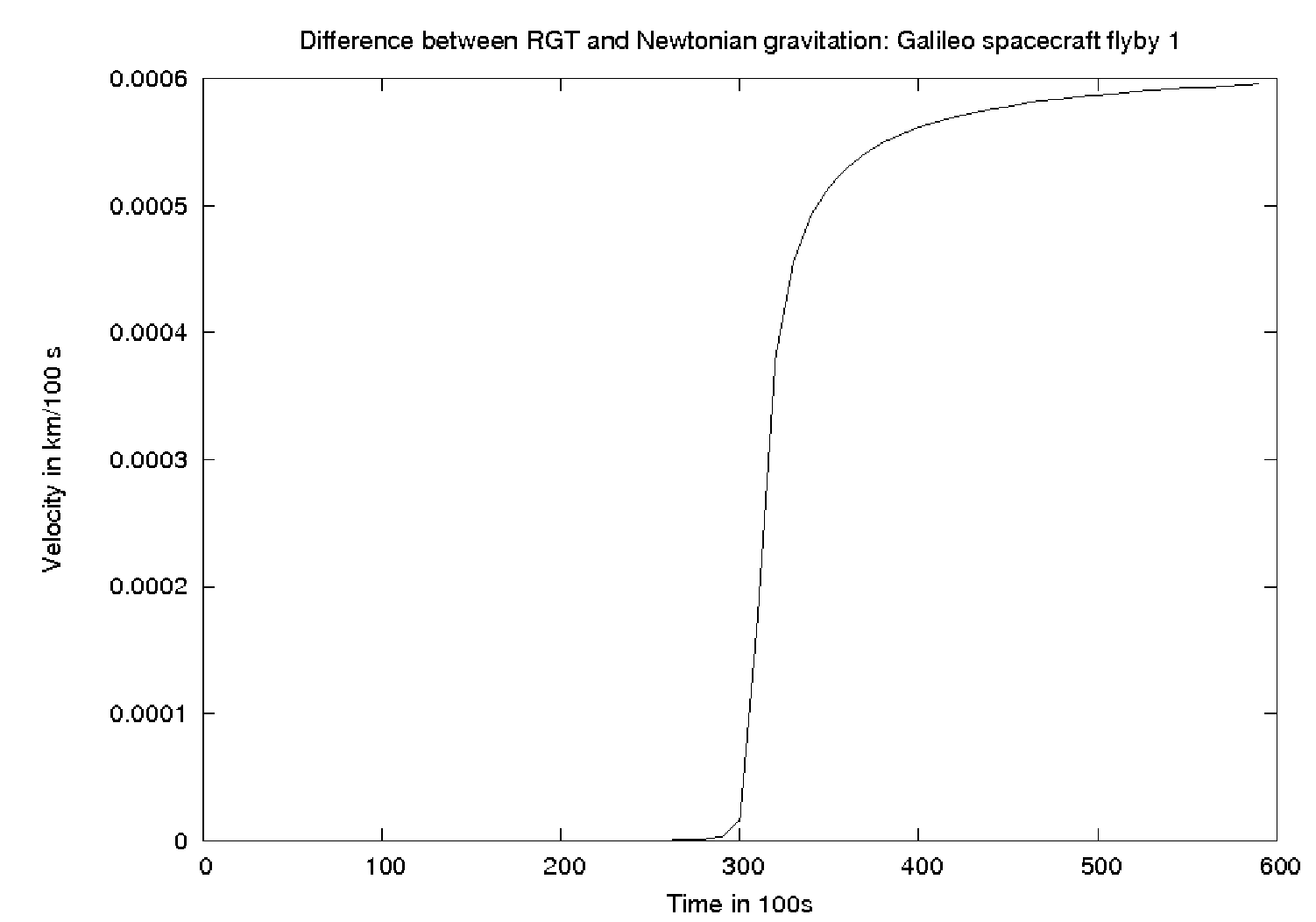}
\caption{
{\bf Galileo.} The difference in the velocity between the solutions obtained using the new velocity-dependent force and the Newtonian force, for the first flyby of the Galileo spacecraft.
The $x$-axis is time (in units of 100~s) and the $y$-axis is difference of (scalar) velocity in units of km per 100~s.}
\label{Galileo}
\end{figure}

The difference of (scalar) velocity for the two solutions (in our chosen units, km per 100 s) is given below for the case of Galileo's first earth flyby (Fig.~\ref{Galileo}) using HORIZONS initial data from about 8 hours before perigee (the starting point is 1990-Dec-08 12:01:00.0000). 
The calculated change in velocity shown in the above figure has the right qualitative features, since most of the gain in velocity is close to perigee. It is also the right order of magnitude, but about 1.5 times larger than the actual change observed for the Galileo spacecraft. 

\begin{figure}[!ht]
\centering
 \includegraphics[width=8 cm]{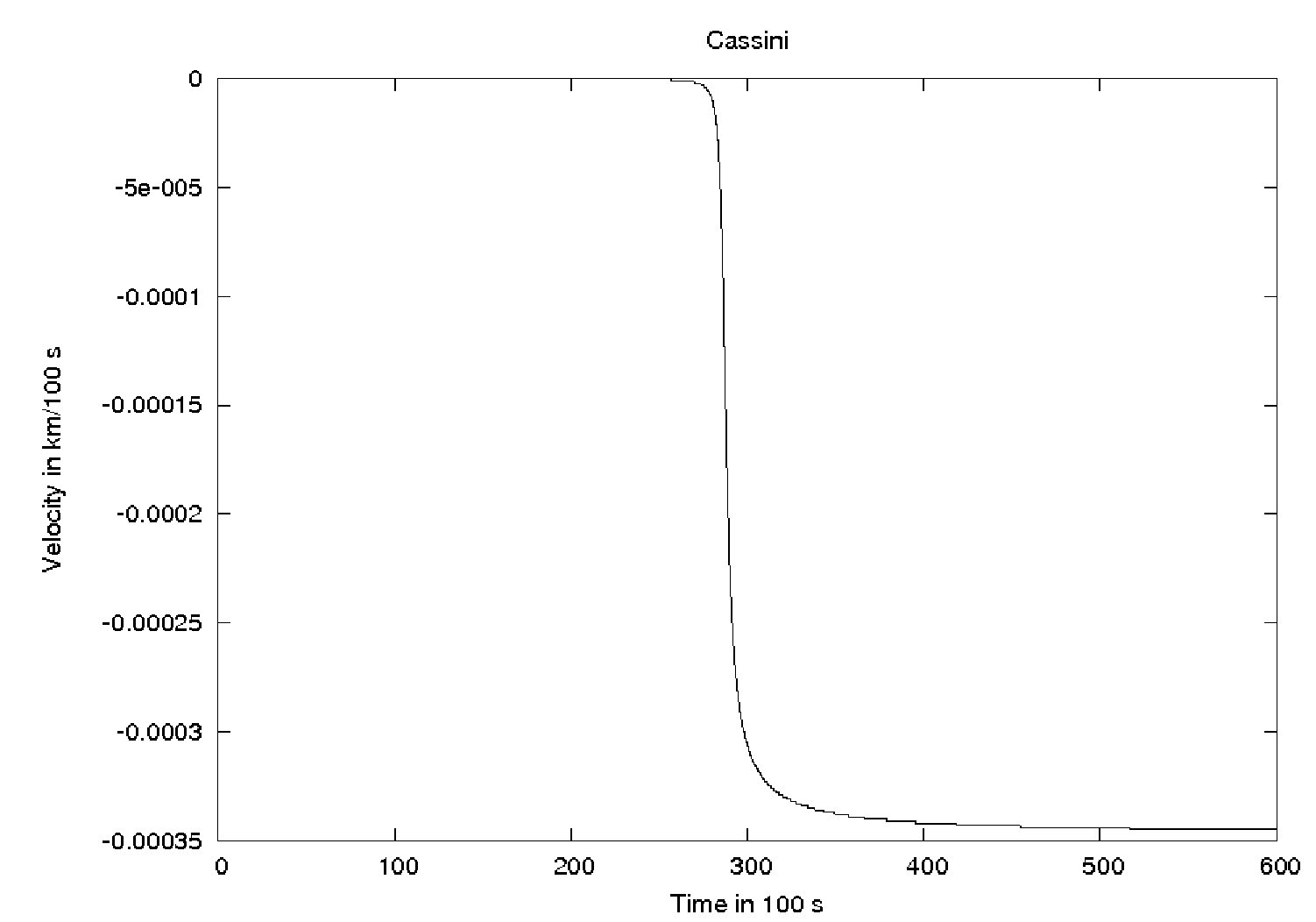}
\caption{
{\bf Cassini.} The difference in the velocity between the solutions obtained using  new velocity-dependent force and the Newtonian force for the earth flyby of the Cassini spacecraft. Same units as before.
The calculated change in velocity is $-3.4$ mm/s compared to the reported change of $-2$ mm/s.}
\label{Cassini}
\end{figure}

There is a more serious discrepancy in the results with Galileo's second flyby. In this case, our calculation leads to +7.6 mm/s gain of velocity instead of the value of $-4.6$ mm/s to $-8$ mm/s reported by Anderson et al., so the calculated result seems off by a factor of $-0.5$ to $-1$. For the NEAR spacecraft our calculation gives a velocity gain of only 3.7 mm/s which is too small by a factor of 3.6. Results for the Cassini spacecraft are shown below (Fig.~\ref{Cassini}). Our calculation yields a change of $-3.4$ mm/s whereas the change reported by Anderson et al.\ is $-2$ mm/s, so the results are again off by a factor of 1.7.

Given the numerous simplifying assumptions we have made, the relatively small discrepancies are very encouraging, and we expect that they will be removed by more refined calculations to be carried out in the future.

\section{Experimental tests}

RGT is not specifically designed to explain galactic rotation curves or exclude dark matter or dark energy or explain the flyby anomaly. Since RGT proceeds from very general theoretical principles, it has numerous other testable consequences.

In the laboratory, RGT can be tested, in principle, as follows. In the conventional Cavendish experiment, suppose the attracting masses are rotated rapidly, in mutually opposite directions. On RGT that would change the deflection of the suspended dumbbell, whereas it should have no effect on the Newtonian theory. Since the expected change is tiny ($\sim\frac{v}{c}$), an unexpected difficulty in calculating it is that the value of $G$ is not known to sufficient accuracy. Hence, an easier way to test RGT might be the following.

The velocity-dependence of gravitation in RGT impinges on current methods of determining the Newtonian gravitational constant $G$. On Newtonian gravitation, the ``static'' and ``time-of-swing'' \cite{NewtonG2} ways of determining $G$ are equivalent, but on RGT one must discriminate between the two.  Of course, the expected difference on RGT is tiny, and there are many confounding factors, so the discrepancies that have actually been found, and even suspected of being due to some fundamental problem \cite{NewtonG}, cannot be put down to this cause, at an acceptable level of confidence. Nevertheless, this discrepancy, between the force-compensation method and the time-of-swing method of determining $G$, is expected on RGT and provides yet another way by which RGT may be tested in the laboratory. More sensitive experiments are anyway required, since  $G$ is the least well-determined of all fundamental constants.

To reiterate, RGT makes no \textit{ad hoc} assumptions, assuming only Lorentz covariance, needed for the measurement of time, so even if RGT were to be experimentally refuted, that would have significant implications across physics.

\section{Conclusions} RGT carries to its logical conclusion the special relativistic correction of Newtonian physics, by making the gravitational force Lorentz covariant. The gravitational force in RGT depends on retarded position and velocity. This correction enables the gravitational many-body problem to be solved in a Lorentz covariant way, using FDEs. Though the velocity dependent effects are small at the level of the solar system, they imply anomalous effects for spacecraft which are inexplicable on Newtonian gravitation or GRT. The differences are large for spiral galaxies, where the velocity drag from a large number of co-rotating stars gives rise to substantial differences from the Newtonian theory. 
RGT is testable in the laboratory. Since RGT makes no \textit{ad hoc} assumptions,  assuming only Lorentz covariance, experimental refutation would have serious implications across physics. RGT necessitates a reappraisal of current methods of determining the Newtonian gravitational constant $G$.

\begin{theacknowledgments}
The author is grateful to the Universiti Sains Malaysia for an Incentive Grant which enabled him to present this paper at Petropolis, Brazil.
\end{theacknowledgments}

\bibliographystyle{aipproc}

\begin{thebibliography}{10}

\bibitem{ckrtitcon}
C.~K. Raju.
\newblock {\em Time: Towards a Consistent Theory}, volume~65 of {\em
  Fundamental Theories of Physics}.
\newblock Kluwer Academic, Dordrecht, 1994.

\bibitem{ckrSced}
C.~K. Raju.
\newblock Time: What is it that it can be measured?
\newblock {\em Science \& Education}, 15\penalty0 (6):\penalty0 537--51, 2006.

\bibitem{Moise}
E.~A. Moise.
\newblock {\em Elementary Geometry from an Advanced Standpoint}.
\newblock Addison Wesley, 1990.

\bibitem{Cultfound}
C.~K. Raju.
\newblock {\em Cultural Foundations of Mathematics}.
\newblock Pearson Logman, 2007.

\bibitem{Hawaii}
C.~K. Raju.
\newblock Computers, mathematics education, and the alternative epistemology of
  the calculus in the {Yuktibh\={a}\d{s}\={a}}.
\newblock {\em Philosophy East and West}, 53\penalty0 (1):\penalty0 325--62,
  2001.

\bibitem{distrib}
C.~K. Raju.
\newblock Distributional matter tensors in relativity.
\newblock In D.~Blair and M.~J. Buckingham, editors, {\em Proceedings of the
  5th Marcel Grossman meeting}, pages 421--23. World Scientific, 1989.

\bibitem{syngeSR}
J.~L. Synge.
\newblock {\em Relativity: the Special Theory}.
\newblock North Holland, Amsterdam, 1956.

\bibitem{ckr3a}
C.~K. Raju.
\newblock Michelson-{M}orley experiment.
\newblock In {\em \textit{Time: Towards a Consistent Theory}}, volume~65 of
  {\em Fundamental Theories of Physics}, chapter~3a, pages 49--58. Kluwer
  Academic, Dordrecht, 1994.

\bibitem{ckrem2bp}
C.~K. Raju.
\newblock The electrodynamic 2-body problem and the origin of quantum
  mechanics.
\newblock {\em Foundations of Physics}, 34\penalty0 (6):\penalty0 937--962,
  2004.

\bibitem{ckrThermo}
C.~K. Raju.
\newblock Thermodynamic time.
\newblock In {\em \textit{Time: Towards a Consistent Theory}}, chapter~4, pages
  79--101. Kluwer Academic, Dordrecht, 1994.

\bibitem{Anderson2008}
John D.~Anderson et~al.
\newblock Anomalous orbital-energy changes observed during spacecraft flybys of
  earth.
\newblock {\em Physical Review Letters}, 100:\penalty0 091102, 2008.

\bibitem{Lammerzahl}
C.~La\"mmerzahl, O.~Preuss, and H.~Dittus.
\newblock Is the physics within the solar system really understood?
\newblock 2006.
\newblock arXiv.gr-qc/0604052.

\bibitem{syngeGR}
J.~L. Synge.
\newblock {\em Relativity: the General Theory}.
\newblock North Holland, Amsterdam, 1960.

\bibitem{ckrJunction}
C.~K. Raju.
\newblock Junction conditions in general relativity.
\newblock {\em J. Phys. A: Math. Gen.}, 15:\penalty0 1785--97, 1982.

\bibitem{RubinA}
V.~C. Rubin, W.~K. Ford, and N.~Thonnard.
\newblock Rotational properties of 21 sc galaxies with a large range of
  luminosities and radii, from ngc 4605 (r=4 kpc) to ugc 2885 (r = 122 kpc).
\newblock {\em Ap. J.}, 238:\penalty0 471--487, 1980.

\bibitem{MilgromA}
M.~Milgrom.
\newblock A modification of {Newtonian} dynamics---implications for galaxies.
\newblock {\em Ap.~J.}, 270:\penalty0 371--383, 1983.

\bibitem{Poincare1906}
H.~Poincar\'e.
\newblock Sur la dynamique de l'\'electron.
\newblock {\em Rend. Circ. Mat. Palermo}, 21:\penalty0 129--176, 1906.

\bibitem{SyngeandGriffith}
J.~L. Synge and B.~A. Griffith.
\newblock {\em Principles of Mechanics}.
\newblock McGraw Hill, 3rd edition, 1959.
\newblock Equation 18.320.

\bibitem{PaTomu}
Suvrat Raju and C.~K. Raju.
\newblock Radiative damping and functional differential equations.
\newblock {\em Modern Physics Letters A}, 26\penalty0 (35):\penalty0 2627--38,
  2011.

\bibitem{NewtonG2}
Jun Luo, Qi~Liu, Liang-Cheng Tu, Cheng-Gang Shao, Lin-Xia Liu, Shan-Qing Yang,
  Qing Li, and Ya-Ting Zhang.
\newblock Determination of the {N}ewtonian gravitational constant ${G}$ with
  time-of-swing method.
\newblock {\em Physical Review Letters}, 102\penalty0 (24):\penalty0 240801(4),
  June 2009.

\bibitem{NewtonG}
Stephen Schlamminger, Eugene Holzschuh, Walter K\"{u}ndig, Frithjof Nolting,
  and J\"{u}rgen Schurr.
\newblock Determination of the gravitational constant.
\newblock In C.~L\"{a}mmerzahl, C.~W.~F. Everitt, and F.~W. Hehl, editors, {\em
  \textit{Gyros, Clocks, Interferometers...: Testing Relativistic Gravity in
  Space}}, pages 15--28. Springer, Berlin, 2001.

\end{thebibliography}

\end{document}